\documentclass[sn-basic]{sn-jnl}

\usepackage{graphicx}%
\usepackage{multirow}%
\usepackage{amsmath,amssymb,amsfonts}%
\usepackage{amsthm}%
\usepackage{mathrsfs}%
\usepackage[title]{appendix}%
\usepackage{xcolor}%
\usepackage{textcomp}%
\usepackage{manyfoot}%
\usepackage{booktabs}%
\usepackage{algorithm}%
\usepackage{algorithmicx}%
\usepackage{algpseudocode}%
\usepackage{listings}%
\usepackage{natbib}
\definecolor{mygreen}{RGB}{0, 185, 118}

\begin{document}

\title[Article Title]{On the analysis of eruptive events with \\ non-radial evolution}


\author*[1]{\fnm{Abril} \sur{Sahade}}\email{asahade@unc.edu.ar}

\author[2]{\fnm{M. Valeria} \sur{Sieyra}}

\author[3]{\fnm{Mariana} \sur{Cécere}}

\affil*[1]{\orgdiv{Heliophysics Science Division}, \orgname{NASA Goddard Space Flight Center}, \orgaddress{\street{8800 Greenbelt Rd}, \city{Greenbelt}, \postcode{20771}, \state{MD}, \country{USA}}}
\affil[2]{\orgdiv{Département d'Astrophysique/AIM}, \orgname{CEA/IRFU, CNRS/INSU, Université Paris-Saclay}, \orgaddress{\street{Université de Paris }, \city{Gif-sur-Yvette}, \postcode{F-91191}, \country{France}}}

\affil[3]{ \orgname{Instituto de Astronomía Teórica y Experimental, CONICET-UNC}, \orgaddress{\street{Laprida 854}, \city{Córdoba}, \postcode{5000}, \state{Córdoba}, \country{Argentina}}}


\abstract{ Coronal mass ejections (CMEs) are major drivers of space weather disturbances, and their deflection from the radial direction critically affects their potential impact on Earth. While the influence of the surrounding magnetic field in guiding CME trajectories is well established, accurately predicting non-radial propagation remains a challenge.

In this work, we introduce and compare recently developed techniques for analyzing the early deflection of eruptive events. We revisit a largely deflected prominence-CME event of 2010 December 16 using an improved tracking framework and a new application of the topological path method. 
Our results suggest the deflection of the eruption is dominated by the channeling of the magnetic field lines. This study offers new physical insight into CME guidance mechanisms and validates the predictive capability of the topological path, highlighting its potential as a diagnostic tool for estimating the propagation direction of strongly deflected events.
}




\maketitle

\section{Introduction}\label{intro}
Solar eruptions are energetic explosive events that inject plasma and magnetic fields into the interplanetary medium. They are typically observed in the lower corona as solar flares and/or prominence eruptions, and later in white-light images of the upper corona as coronal mass ejections (CMEs). When directed toward Earth, CMEs can trigger severe geomagnetic storms capable of disrupting global communication and navigation systems, damaging satellites, and causing large-scale power grid failures. As the number of missions operating throughout the heliosphere continues to grow, accurately predicting the evolution of CMEs in the interplanetary medium has become more critical than ever.

A key factor in determining whether an eruptive event will impact Earth or a spacecraft is its direction of propagation. Recent statistical studies have shown that CMEs frequently exhibit some degree of deflection from their source region, rather than propagating strictly radially \citep[see e.g.,][]{Majumdar2020,Majumdar2023}. These deflections can be significant \citep[e.g.,][]{Isavnin2014, Sieyra2020}, sometimes resulting in false space weather alerts \citep[see e.g.,][for an analysis of the 2014 January 7 event]{Mostl2015, Mays2015}. It is well established that the magnetic fields surrounding the source region play a critical role in governing CME deflection. This influence has been extensively studied through both observational \citep[e.g.,][]{Kilpua2009, Kay2017, Cecere2020, Sahade2023} and numerical \citep[e.g.,][]{Sahade2020, Sahade2021, Sahade2022, Ben-Nun2023} approaches. A recent review by \citet{Cecere2023} summarizes the current understanding of the key factors contributing to early-stage CME deflections.

Building on numerical simulations of magnetic flux rope (MFR) deflections in various magnetic environments, and comparisons with observational case studies \citep{Sahade2023PhDT}, a new method has been proposed to estimate CME deflections \citep{Sahade2025}. This method introduces the `topological path,' defined as a curve connecting the apexes of closed and open magnetic field lines above the source region. This path approximates the expected CME trajectory, assuming it is guided (``channeled'') by the surrounding magnetic field. The study also defines the `gradient path,' which follows the direction of minimal magnetic energy from the source region, estimating the deflections driven by the magnetic energy gradient. \citet{Sahade2025} applied these methods to eight prominence-CME events, demonstrating that the topological path provides a promising tool for estimating propagation direction in both radially directed and strongly deflected events. A key advantage of these methods is that they rely only on the location of the source region and the overlying magnetic field, making them suitable for forecasting purposes.

The present study introduces an improved framework to quantify the role and mechanism of the ambient magnetic field in guiding CME trajectories.
We revisit an event previously analyzed by \citet{Sieyra2020}, using newly developed tracking tools \citep{Sahade2023,Sahade2025}. This prominence-CME event was simultaneously observed from multiple vantage points of view and exhibited a large deflection ($>20^\circ$) during its early evolution. The CME propagated toward the heliospheric current sheet (HCS) showing alignment with the gradient of the magnetic energy. This event provides an ideal test case for evaluating the diagnostic potential of the gradient and topological path concepts.
Crucially, we show that although the CME trajectory aligns with the magnetic energy gradient, the physical mechanism driving the early deflection is more likely governed by the magnetic field topology. Our work offers a new perspective on CME guidance mechanisms and contributes to improving predictive models for space weather forecasting.
In Section~\ref{met}, we describe the enhanced tracking techniques and introduce the refined analysis of magnetic paths that support this interpretation. Results and conclusions are presented in Sections~\ref{res} and~\ref{con}, respectively.

\section{Methodology}\label{met}
\subsection{Data and tracking}\label{ss:track}

From the events analyzed by \citet{Sieyra2020}, we selected the 2010 December 16 event due to its excellent observational coverage throughout its evolution, which allowed us to track the eruption up to approximately $8\,R_\odot$ from multiple viewpoints. The erupting prominence-CME was observed simultaneously from Earth by the \textit{Solar Dynamics Observatory} \citep[SDO;][]{SDO_2012SoPh} and the \textit{Large Angle and Spectrometric Coronagraph Experiment} \citep[LASCO;][]{LASCO_1995SoPh}, as well as by the twin spacecraft of the \textit{Solar TErrestrial RElations Observatory} \citep[STEREO;][]{STEREO_2008}—STEREO-Ahead (STA) and STEREO-Behind (STB)— which were positioned near quadrature during this period.
In addition, the location of the source region enabled the use of high-quality photospheric magnetograms from the \textit{Helioseismic and Magnetic Imager} \citep[HMI;][]{HMI2012}. These updated magnetograms provide more reliable boundary conditions for reconstructing the ambient magnetic field surrounding the source region. To model the global structure of the coronal field, we employed the Potential Field Source Surface (PFSS) model \citep{PFSS_2003SoPh}, which estimates the magnetic field configuration up to $2.5,R_\odot$.

\begin{figure}
    \centering
    \includegraphics[width=0.9\textwidth]{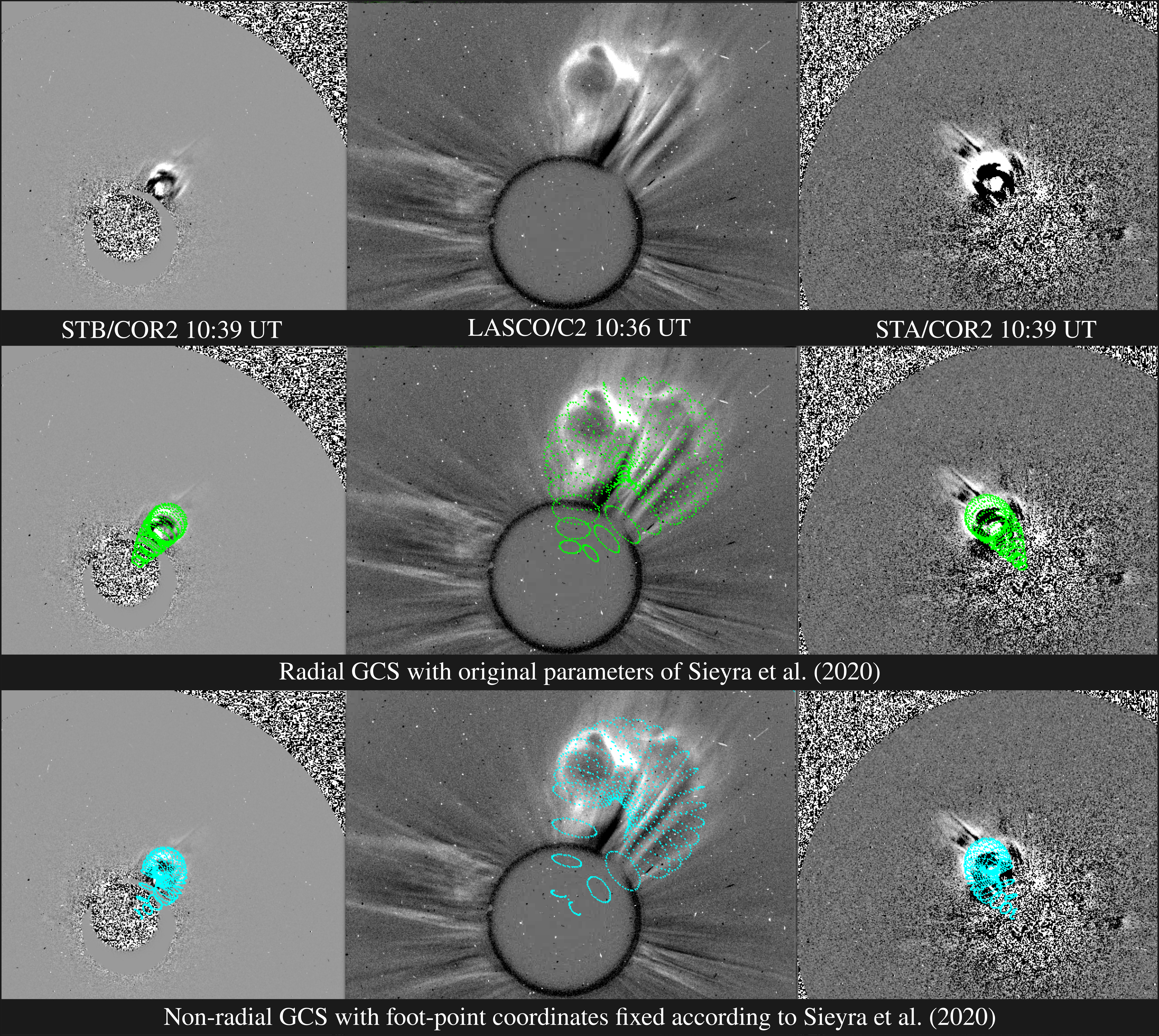} 
    \caption{Upper panel: Running difference of the white-light images from STA, LASCO and STB perspectives of the 2010 December 16 CME at 10:36 UT. Middle panel: GCS wireframe (green lines) projected over the coronograph images showed in upper panel. Lower panel: non-radial GCS wireframe (cyan lines) projected on STA, LASCO and STB FOV.}
    \label{fig:GCS}
\end{figure}

To characterize the evolution of the prominence material, we use the tie-pointing (TP) reconstruction technique \citep[see e.g.,][]{Inhester2006, Mierla2008, mierla2009} applied to EUV images from SDO/AIA and wavelet-enhanced images \citep{Stenborg2008} from the \textit{Extreme-Ultraviolet Imager} \citep[STEREO/EUVI;][]{SECCHI_2008}. TP is a stereoscopic method that determines the three-dimensional position of an object (in this case, a plasma parcel) based on its projection in a pair of images. The key challenge lies in identifying the same feature in both images, i.e. matching a feature in image 1 and its corresponding projection in image 2 \citep[see][for details on the challenges and sources of error]{Inhester2006}.
In this case, we recalculate the position of the prominence using the \texttt{scc\_measure3} routine \citep{scc_measure3}, which derives the 3D coordinates of corresponding pixels using images from three different vantage points. As in the standard TP method, the process begins with selecting the same feature in two images to determine its 3D location. This position is then projected onto the third image, which serves as a validation step to confirm or reject the initial selection \citep[see Appendix of][]{Sahade2025}. The third viewpoint acts as a constraint, helping to resolve ambiguities and eliminate ghost features (artifacts) not associated with the actual material being tracked.
Although the angular separation between STEREO-A and STEREO-B was approximately $173^\circ$, resulting in highly similar projected views, the inclusion of the third image significantly facilitated and accelerated the identification of matching features.

\begin{figure}
    \centering
    \includegraphics[width=0.9\textwidth]{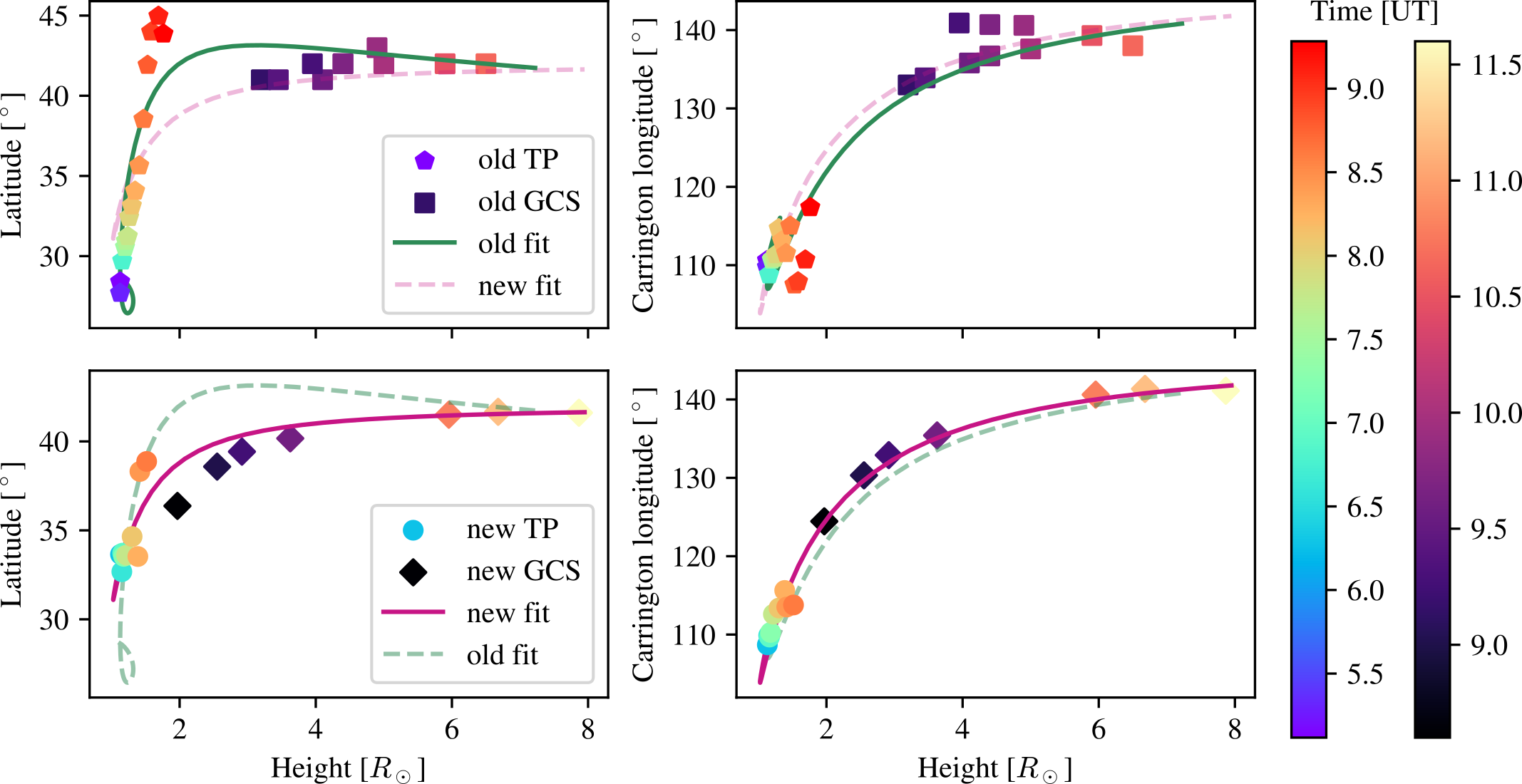} 
    \caption{Upper panels: Coordinates from the \citet{Sieyra2020} reconstruction. The pentagons indicate the position obtained by TP between STA and AIA 304{\AA} images. The colorscale indicates the time (UT) of the observation as shown in the first colorbar. The squares indicate the apex coordinates of the radial GCS obtained between LASCO C2, with COR1 and COR2 FOV. The colorscale indicates the time (UT) of the observation as shown in the second colorbar. The solid green line shows the spline fitting for the coordinates over time. For comparison, the dashed pink line shows the corresponding fitting for the new coordinates. Lower panel: Coordinates obtained with the new techniques. The circles indicate the position obtained by TP between STA, AIA, and STB 304{\AA} images. The diamonds indicate the apex coordinates of the non-radial GCS obtained between LASCO C2, COR1 and COR2 FOV. Both colorscales are the same as upper panel. The solid pink line shows the spline fitting for the new coordinates over time. For comparison, the dashed green line shows the corresponding fitting for the old coordinates.}
    \label{fig:coords}
\end{figure}

To track the evolution of the ejected material in coronagraph images, we fitted the CME using the Graduated Cylindrical Shell (GCS) model \citep{Thernisien2006, Thernisien2009}. Given the known deflection of the CME, we applied the non-radial version of the GCS model, which includes an additional degree of freedom: the tilt of the CME axis relative to the radial plane (defined by the solar center and the CME footpoints). This approach allows the CME front to vary in both latitude and longitude, while keeping the footpoints fixed at the source region. The fitting was performed using the \texttt{rtcloudwidget} routine available in \textit{SolarSoft}.  The non-radial tilt is controlled by the \texttt{Ne Tilt} slider under the \texttt{Ne Shift} tab, which is part of the standard GCS fitting interface but is not commonly used in CME reconstructions. 

Figure~\ref{fig:GCS} illustrates this fitting process. The first panel shows the 2010 December 16 CME as observed around 10:36 UT in the field of view (FOV) of STA/COR2, LASCO/C2, and STB/COR2. The second panel displays the original GCS wireframe (green lines) from \citet{Sieyra2020}, with the following parameters: Carrington longitude $138^\circ$, latitude $42^\circ$, height $6.6,R_\odot$, tilt angle $-30^\circ$, half-angle $18^\circ$, and aspect ratio $0.27$. The third panel shows the updated non-radial GCS fit (cyan lines), with parameters: Carrington longitude $110^\circ$, latitude $29^\circ$, height $6.7,R_\odot$, tilt angle $-55^\circ$, half-angle $18^\circ$, aspect ratio $0.2$, and a non-radial tilt of $33^\circ$. In this non-radial configuration, the longitude, latitude, and tilt angle remain constant throughout the CME’s evolution within the coronagraph FOV, and these coordinates correspond to the source region location reported by \citet{Sieyra2020}. Due to the non-radial tilt, the apex of the CME does not share the same coordinates as the footpoints. 
Each reconstruction (old and new) provides a consistent fit over time, which helps constrain the parameter space and reduce the associated uncertainties \citep{Verbeke2023}. For this event and dataset, changes of approximately $1^\circ$ in the angular parameters still fit the CME front. Therefore, we consider this to be the associated error. Taking into account the source region location and tilt, we aim to minimize projection-related ambiguities and more accurately represent the front evolution, particularly in the context of deflected propagation.
The most significant difference between the original and updated reconstructions is the tilt angle, whose implications for space weather forecasting will be discussed in Section~\ref{con}. While the difference between reconstructions ($25^\circ$) falls within the expected range \citep[as indicated by previous studies, e.g.,][]{Verbeke2023,Kay2024}, it is important to note that this difference should not be interpreted as an associated error. For example, if we take the parameters from the new fitting and vary the tilt angle back to $-30^\circ$ (the value from the original reconstruction), the GCS model no longer fits the CME. This highlights the need to account for the non-radial evolution of CMEs in order to accurately reconstruct the orientation of the CME flux rope.

Figure~\ref{fig:coords} presents the evolution of the prominence-CME latitude and longitude during its ascent. The upper panels show the previous tracking: the tie-pointed (TP) prominence coordinates (pentagon markers) and the GCS apex positions (square markers) from \citet{Sieyra2020}. The lower panels show the new tracking results: prominence positions (circle markers) and non-radial GCS apex positions (diamond markers). From the 3D Cartesian coordinates, we modeled the eruption path using a cubic spline as a function of time (with the \texttt{scipy.interpolate.UnivariateSpline} function). The trajectory fitting was done in Cartesian space, where the variations in the three coordinates are comparable ($dx = dy = dz$), but results are presented in spherical coordinates for clarity.
The green line (solid in the upper panels, dashed in the lower) represents the eruption path from the original tracking, while the pink line (dashed in the upper panels, solid in the lower) corresponds to the new tracking. The colorbars indicate observation times in EUV (rainbow scale) and white-light (magma scale) images. In the right panels of Fig.~\ref{fig:coords}, we see that both reconstructions yield similar longitude profiles. However, the new reconstruction exhibits reduced scatter and a smoother transition between TP and GCS coordinates, as expected given the added constraint from the third TP viewpoint and the improved fit to the CME shape in its early stages.
The largest difference between the reconstructions appears in the latitude (left panels). In the original tracking, the prominence position reached up to $45^\circ$, whereas the new reconstruction shows a lower maximum latitude. Although both methods ultimately converge to similar final CME positions, there is a small discrepancy in the times at which those positions are recorded (comparing the square markers in the upper panel with the diamond markers in the lower panel). This difference arises from the use of different GCS models to reproduce the CME projected appearance.

\subsection{Magnetic influence}

Following the methodology presented in \citet{Sahade2025} we calculate two `magnetic paths' originating from the source region of the event. The `gradient path,' defined from the initial position by following the ascending direction of minimal magnetic energy, represents the trajectory that an eruption would follow if influenced solely by the gradient of the magnetic field.  Similarly, the `topological path,' which connects the apexes of the overlying field lines, represents the trajectory of an eruption that moves ``channelized'' along the magnetic field lines. These paths help us understand whether an eruptive event is primarily deflected by magnetic pressure or by the ``shape'' of the magnetic field lines. Therefore, we compare the final positions of the magnetic paths, originating from the source region of the eruptive event, with the eruptive path (computed as explained in the previous section) up to the source surface ($2.5\,R_\odot$). The algorithms to calculate each path are available at \burl{https://github.com/asahade/Events/tree/main/20101216}. 

\begin{figure}
    \centering
    \includegraphics[width=0.96\textwidth]{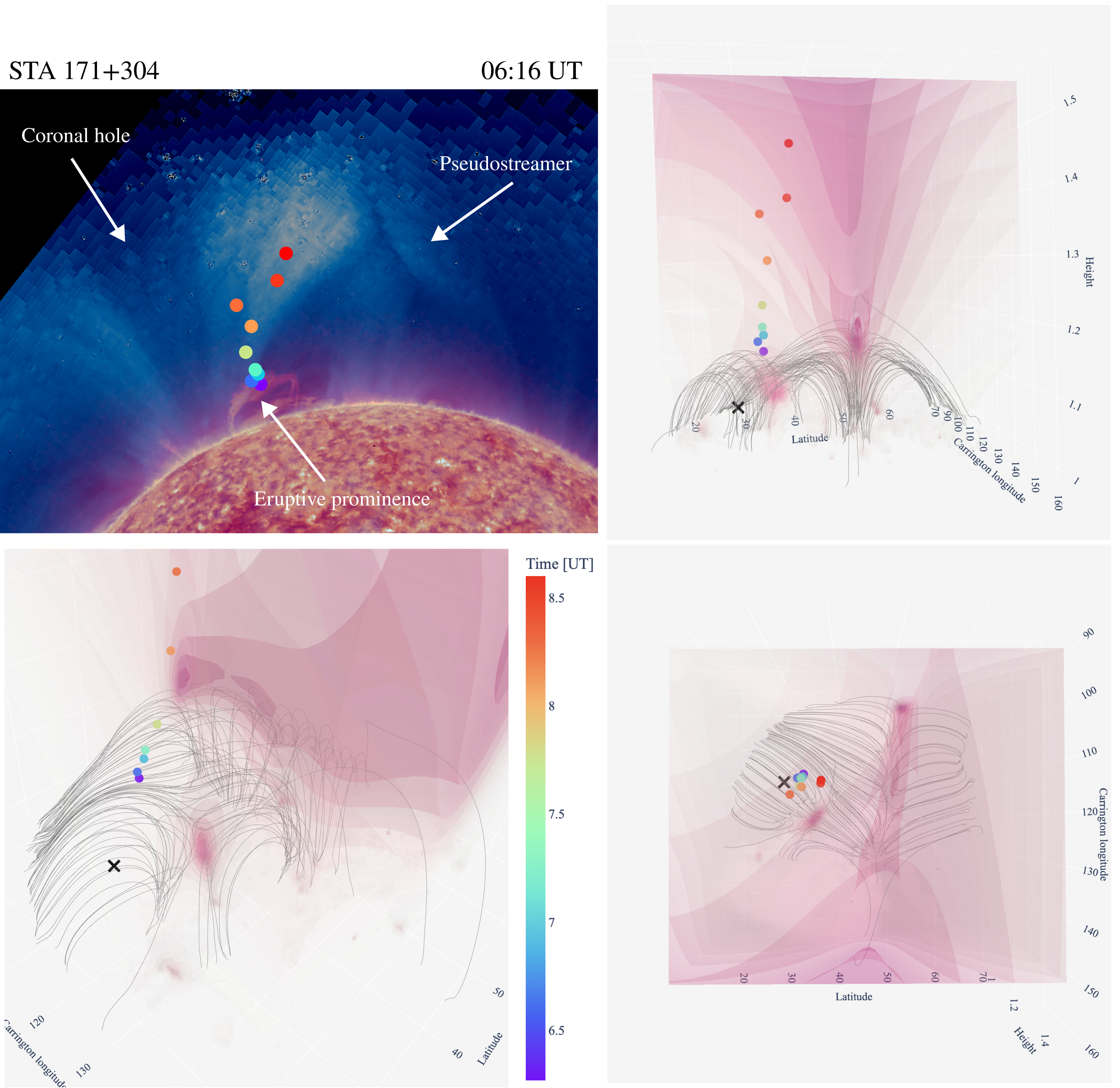} 
    \caption{ Trajectory of the eruption and its magnetic context. \textit{Top left panel:} Pre-eruptive composite image (171 \AA{} and 304 \AA{}) of the eruptive prominence from the STA perspective. Nearby magnetic structures, such as a coronal hole and a pseudostreamer, are indicated over the solar limb.
    \textit{Other panels:}
    Three-dimensional views of the eruption trajectory and its surrounding magnetic environment up to a height of $1.5\,R_\odot$. Gray lines represent closed magnetic field lines, showing the arcades overlying and surrounding the source region of the eruption (marked by the black cross at coordinates $(110^\circ, 29^\circ)$). Colored dots represent the tracked position of the erupting prominence over time, with the corresponding time (in hours) indicated by the color bar. Pink contours show isosurfaces of magnetic energy, with darker shades representing regions of lower magnetic energy. All magnetic field structures and isocontours are based on the PFSS extrapolation, and 3D prominence positions were obtained using the tie-pointing tracking method described in Section~\ref{ss:track}.
    }
    \label{fig:magenv}
\end{figure}

The eruptive prominence was originally located under the lobe of a complex pseudostreamer region. The simplest pseudostreamers are formed from an embedded-bipole topology surrounded by unipolar fluxes \citep[e.g.,][]{Raouafi2016,Mason2021,Wyper2021}, forming arcades capped by magnetic field lines from coronal holes of the same polarity. These field lines converge at a central spine with a null point at its base \citep[e.g.,][]{Sahade2022}. However, as magnetic polarities become mixed, the formation of new arcade systems can result in a more complex topology with local null points and corridors. Figure~\ref{fig:magenv} shows the tracked position of the prominence and its magnetic context. The first panel shows the STA view of the pre-eruptive configuration. A coronal hole and pseudostreamer are identified in the limb of the composite ($171+304$ EUVI) image, within the prominence and its tracking. With the magnetic field obtained from the PFSS model, the rest of the panels show different views of the overlying magnetic field under which the prominence erupted. The gray lines represent closed magnetic field arcades surrounded by a negative polarity coronal hole (see blue lines in Fig.~\ref{fig:paths}). Pink contours indicate surfaces of equal magnetic energy, with darker shades corresponding to regions of lower magnetic energy. At low heights, two null regions can be identified: one at the top of the pseudostreamer (dark pink area at latitude $\sim 50^\circ$ and height $\sim 1.2,R_\odot$), and a secondary null embedded in the southern lobe of the pseudostreamer (dark pink at latitude $\sim 30^\circ$ and height $\sim 1.1,R_\odot$). The source region (black cross), located at $(110^\circ,29^\circ)$ as reported in \citet{Sieyra2020}, was embedded in the southwestern lobe of the pseudostreamer, where mixed polarities created a complex system of triple arcades an the secondary null point (see middle and right panels of the figure). Within this magnetic context, the prominence erupted, following the path indicated by the colored dots.

\section{Results and Discussion}\label{res}

\begin{figure}
    \centering
    \includegraphics[width=0.96\textwidth]{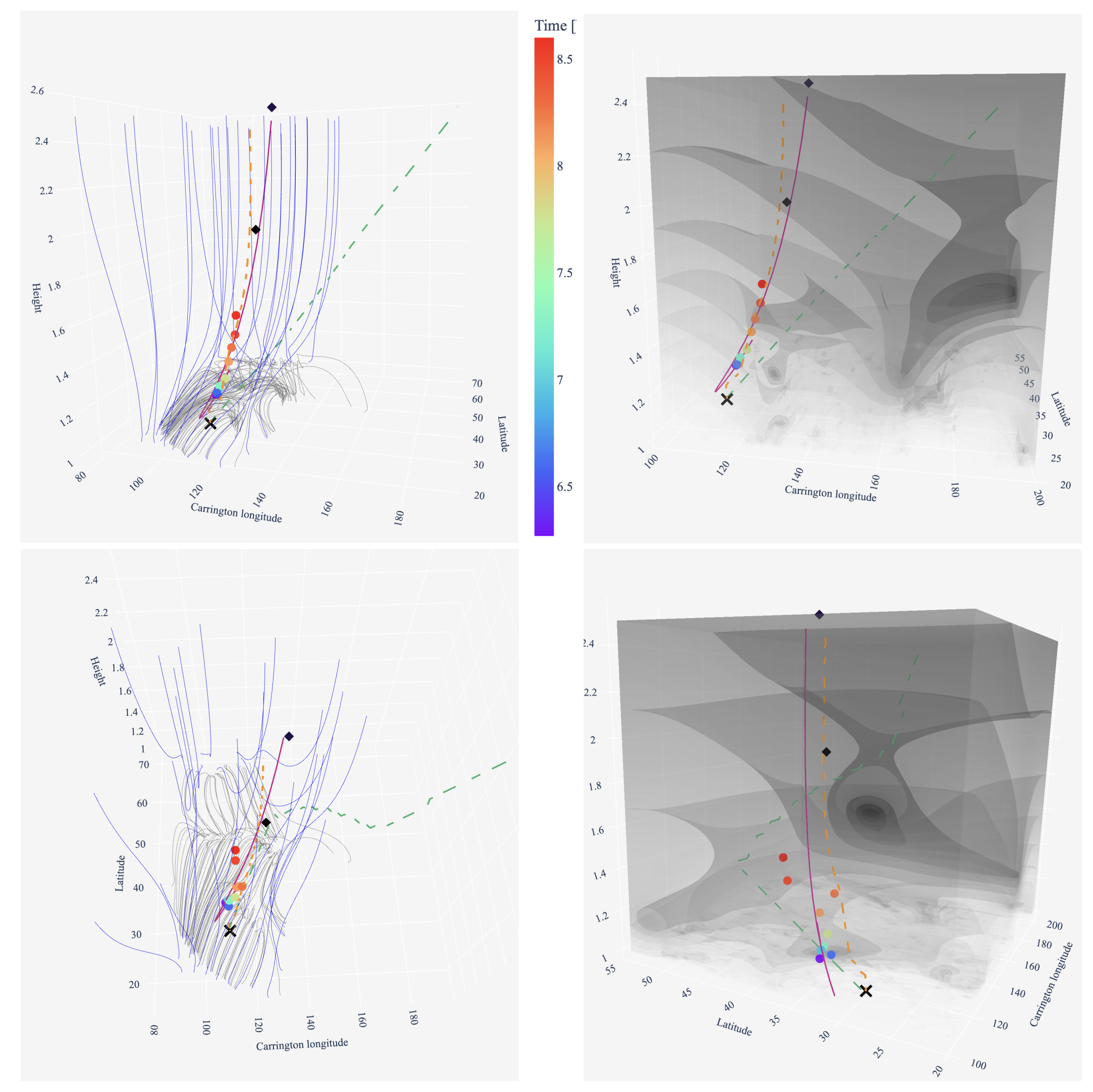} 
    \caption{Views of the eruptive path (solid pink line), the topological path (dashed orange line), and the gradient path (dashed green line) in the magnetic context up to $2.5\,R_\odot$. The colored dots indicate the position of the prominence as evolved in time (indicated by the colorbar), and diamonds the position of the CME under $2.6\,R_\odot$. Left panels show closed (gray lines) and open (blue lines) magnetic field lines overlying and surrounding the source region (black cross) of the eruption.  Right panels show the magnetic energy isocontours in gray scale, with darker grey showing the lower energy regions.}
    \label{fig:paths}
\end{figure}

Starting from the source region position, we computed the topological path by following the tops of the overlying arcades as previously described, and the gradient path by tracing the direction of decreasing magnetic energy. The specific parameters and codes used to generate these results can be found at \burl{https://github.com/asahade/Events/tree/main/20101216}. 
Figure~\ref{fig:paths} presents different views of the resulting paths within the magnetic environment. The left panels display magnetic field lines to provide context for the topological path (dashed orange line). This path follows the apex of the overlying arcades below $\sim1.2,R_\odot$ and then transitions to the open field line topology of the southern coronal hole. The right panels show magnetic energy isocontours, providing context for the gradient path (dashed green line). This path is initially directed toward the secondary null point described earlier, then toward the pseudostreamer null point, and eventually toward a global minimum near the heliospheric current sheet (HCS). When compared to the eruptive path (solid pink lines and rainbow scale markers), the gradient path —representing a prominence influenced solely by magnetic energy— exhibits stronger deflections and is guided toward more localized energy minima. In contrast, the topological path —representing a prominence channeled along magnetic field lines— follows a trajectory that closely matches the actual eruptive path.

\begin{figure}
    \centering
    \includegraphics[width=0.9\textwidth]{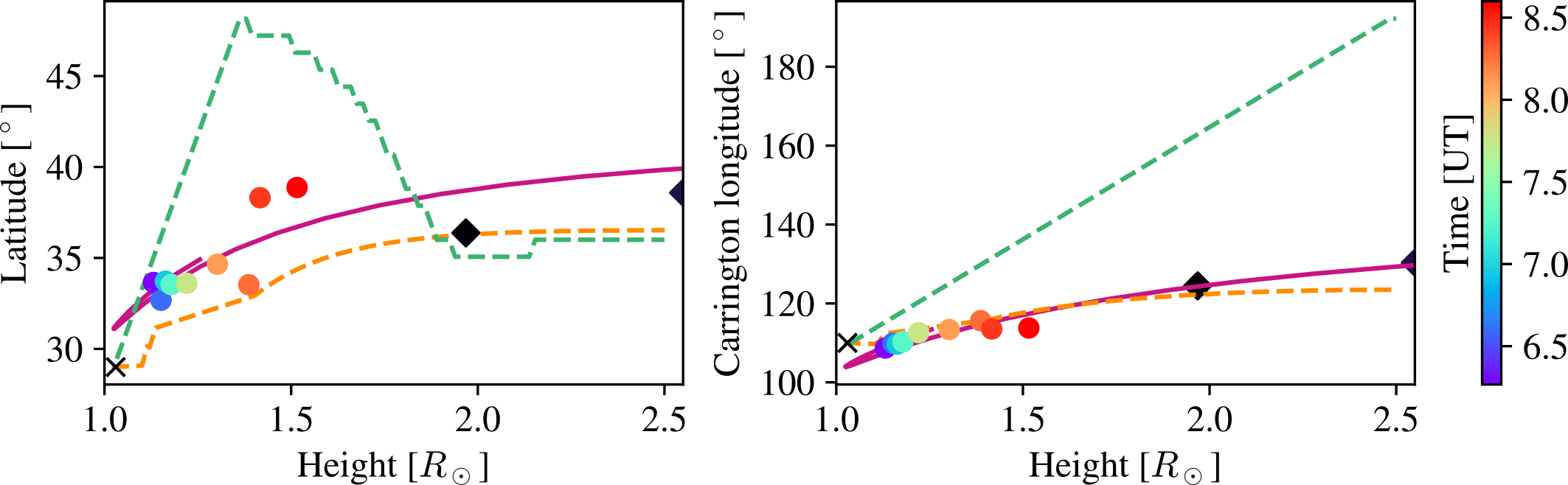} 
    \caption{Comparison between the eruptive and magnetic paths. As in Fig.~\ref{fig:coords}, the colored dots indicate the position of the prominence obtained by TP and the diamonds indicate the apex coordinates of the non-radial GCS.  The solid pink line shows the spline fitting for the eruptive path. The dashed green line shows the gradient path coordinates and the dashed orange line shows the topological path ones.}
    \label{fig:path_coords}
\end{figure}

The magnetic paths proposed here represent a simplified scenario in which only one factor influences the deflection of the eruption. Neither the internal dynamics of the eruptive magnetic system nor its interaction with the surrounding magnetic field is taken into account. Nevertheless, this approach is useful for two main purposes: (1) to determine the dominant factor governing the trajectory of eruptive events, and (2) to predict the direction of propagation based on the location of the source region.
To address the first point, we compare the latitudinal and longitudinal deflections of each magnetic path with the actual trajectory of the eruption, as shown in Figure~\ref{fig:path_coords}. As seen in Figure~\ref{fig:paths}, the gradient path (dashed green line) exhibits greater deflection than the prominence in both latitude and longitude, while the topological path remains closer to the observed trajectory. This suggests that channeling along magnetic field lines had played a major role in the 2010 December 16 eruption evolution. Furthermore, examining discrepancies below $1.5,R_\odot$ in the latitudinal deflection (left panel), it is possible that the local null point also had contributed to deflection toward higher latitudes.

\begin{figure}
    \centering
    \includegraphics[width=0.9\textwidth]{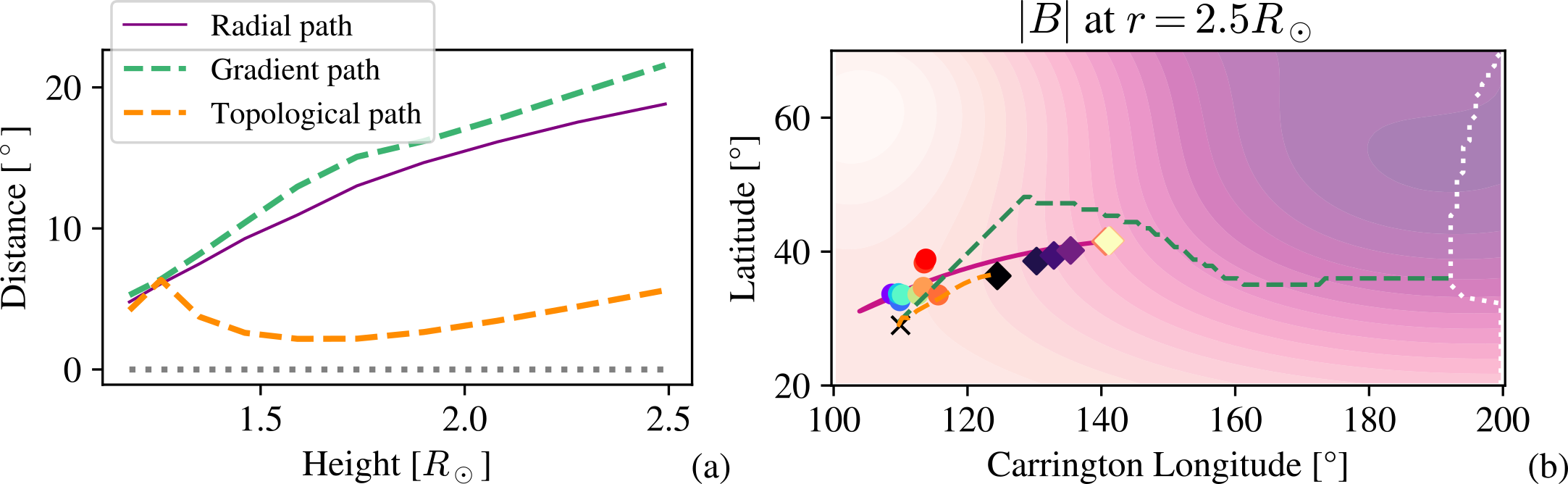} 
    \caption{Comparison between the eruptive and magnetic paths. As in Fig.~\ref{fig:coords}, the colored dots indicate the position of the prominence obtained by TP and the diamonds indicate the apex coordinates of the non-radial GCS.  The solid pink line shows the spline fitting for the eruptive path. The dashed green line shows the gradient path coordinates and the dashed orange line shows the topological path ones.}
    \label{fig:forecast}
\end{figure}

We also quantify the total angular distance between the actual trajectory and the magnetic paths, which serves as a metric for how well each path estimates the eruption's direction. Figure~\ref{fig:forecast}(a) shows the angular distances of the gradient path (dashed green line), the topological path (dashed orange line), and a purely radial path (violet line) from the observed trajectory. The smaller the angular distance (closer to $0^\circ$), the better the estimation. The topological path remains within $10^\circ$ of the eruptive path, indicating it provides a more accurate prediction than assuming a radial propagation. In contrast, the gradient path shows a large and increasing deviation from the eruptive path, making it unsuitable for directional forecasting.
We computed the magnetic paths up to $2.5,R_\odot$, consistent with the PFSS model used to derive the coronal magnetic field, which assumes a source surface at that height. However, from Figure~\ref{fig:coords}, we note that the CME deflection stabilizes around $\sim4,R_\odot$. Figure~\ref{fig:forecast}(b) shows a latitude–longitude map of the paths overlaid on the magnetic field magnitude $|B|$ at $2.5,R_\odot$. Darker purple regions correspond to lower magnetic energy, and the white dotted line marks the position of the heliospheric current sheet (HCS). The diamonds indicate the position of the CME, with all but the black one corresponding to heights above $2.5,R_\odot$.
We observe that from this position, the CME continues traveling in the direction of the magnetic energy gradient—toward darker $|B|$ contours (not along the gradient path -green dashed line, which lies below). This result is in agreement with the magnetic energy density gradient model \citep{Shen2011,Gui2011}, which posits that the net force acting on the CME is dominated by magnetic pressure. Beyond $2.5,R_\odot$, where radial field lines are assumed, magnetic tension is expected to be symmetrically distributed around the CME, and thus largely canceled out.

\section{Conclusions}\label{con}

The dynamics underlying solar eruptions are inherently complex. Advancing our understanding, and our space weather forecasting capabilities, requires a combined effort involving numerical modeling, theoretical frameworks, and detailed analysis of observations. In this work, we apply novel techniques and methods \citep[presented in][]{Sahade2023,Sahade2025} to study the evolution of solar eruptive events with non-radial propagation, focusing on how the surrounding magnetic environment influences this evolution.
By revisiting a previously analyzed event using our enhanced tracking and modeling techniques, we were able to directly assess and validate the improvements in methodology, demonstrating smoother trajectory reconstruction, more confident feature identification, and greater physical insight into the eruption dynamics.

We analyzed the prominence–CME eruption of 2010 December 16 \citep[previously studied by][]{Sieyra2020}. The main advantage of applying a tie-pointing technique with three viewpoints \citep{scc_measure3} lies in its ability to reliably match features across multiple images. The additional constraint improves feature identification and reduces the need for time-consuming visual inspection of small, similar structures in two-image comparisons. The final eruptive position obtained with the new tracking is in very good agreement with previous results \citep{Sieyra2020}, but the improved smoothness and consistency of the new trajectory lend greater confidence to the technique. For the GCS reconstruction, we introduced a new free parameter, the non-radial tilt, which led to a new set of valid parameters that reproduce the CME front evolution. This approach has the added advantage of incorporating source region information to reduce projection ambiguities. The most significant difference in the GCS reconstruction was the tilt angle, a key parameter for forecasting geoeffectiveness, as it affects the $B_z$ component and the extent of magnetic erosion \citep{Kilpua2017}.  The combination of enhanced three-viewpoint tracking and non-radial GCS reconstructions demonstrated here offers a robust framework for future studies and operational tools in space weather forecasting.

To assess the role of the magnetic environment, we compute two theoretical trajectories: one governed solely by the magnetic energy gradient (the gradient path), and the other fully channeled along magnetic field lines (the topological path). Consistent with the findings of \citet{Sahade2025}, we find that the actual prominence trajectory closely follows the topological path, suggesting that deflection during the early evolution of the eruption is primarily driven by magnetic field line channeling. Furthermore, considering that the error in this estimation is smaller than the deviation from a radial trajectory, the topological path emerges as a more accurate predictor of propagation direction.

With the continuing advances in observational capabilities provided by current and upcoming solar missions (e.g., Solar Orbiter, Parker Solar Probe, PUNCH, Vigil), our ability to view eruptions from multiple perspectives is increasing. This offers an unprecedented opportunity to deepen our understanding of solar activity. At the same time, it presents the challenge of integrating diverse datasets to improve predictive models. We hope that the results presented in this work contribute meaningfully to that ongoing effort.

\backmatter




\section*{Declarations}

\begin{itemize}
\item \textbf{Funding:} AS was supported by an appointment to the NASA Postdoctoral Program at the NASA Goddard Space Flight Center, administered by Oak Ridge Associated Universities under contract with NASA. MVS acknowledges support from the French Research Agency grant ANR STORMGENESIS \#ANR-22-CE31-0013-01. MC is member of the Carrera del Investigador Científico (CONICET). MC acknowledges support from SeCyT, UNC. Proyecto Formar 2023 Nº 33820230100116CB, and from the DynaSun project, which has received funding under the Horizon Europe programme of the European Union, grant agreement no. 101131534. Views and opinions expressed are however those of the author(s) only and do not necessarily reflect those of the European Union and therefore the European Union cannot be held responsible for them.

\item \textbf{Competing interests:} The authors have no relevant financial or non-financial interests to disclose.
\item \textbf{Author contribution:} Conceptualization: Abril Sahade, M. Valeria Sieyra; Methodology, formal analysis and investigation: Abril Sahade; Material preparation, data collection: Abril Sahade, Mariana Cécere; Writing - original draft preparation: Abril Sahade. All authors read and approved the final manuscript.
\end{itemize}

\noindent


\bibliography{biblio}  

\IfFileExists{\jobname.bbl}{} {\typeout{}
\typeout{****************************************************}
\typeout{****************************************************}
\typeout{** Please run "bibtex \jobname" to obtain} \typeout{**
the bibliography and then re-run LaTeX} \typeout{** twice to fix
the references !}
\typeout{****************************************************}
\typeout{****************************************************}
\typeout{}}

\end{document}